\documentclass[manuscript]{acmart}
\AtBeginDocument{%
  }
\usepackage{listings}
\usepackage{amsmath}
\usepackage{braket}
\definecolor{comment-text-color}{rgb}{0,0.8,0.6}
\usepackage{listings}
\usepackage{xcolor}
\lstdefinelanguage{Julia}{
  keywords={function, end, if, else, for, while, return, struct},
  sensitive=true,
  comment=[l]{\#},
  string=[b]"
}
\definecolor{codegray}{rgb}{0.95,0.95,0.95}   
\definecolor{codeblue}{rgb}{0.25,0.35,0.75}   
\definecolor{codegreen}{rgb}{0,0.6,0}         
\definecolor{codered}{rgb}{0.60,0,0}          
\definecolor{codeviolet}{rgb}{0.50,0.10,0.60} 
\definecolor{codeorange}{rgb}{0.80,0.35,0.00} 
\definecolor{linegray}{rgb}{0.50,0.50,0.50}   

\lstdefinestyle{juliastyle}{
    language=Julia,
    backgroundcolor=\color{codegray},
    basicstyle=\ttfamily\footnotesize,
    keywordstyle=\color{codeblue}\bfseries,        
    stringstyle=\color{codered},                   
    commentstyle=\color{codegreen}\itshape,        
    alsoletter=@_,                                 
    morekeywords=[2]{NTuple,Cdouble,Float64,Int64},
    keywordstyle=[2]\color{codeviolet}\bfseries,
    morekeywords=[3]{@ccallable},                  
    keywordstyle=[3]\color{codeorange}\bfseries,
    emph={computeMidpointMPSEntropy,compute_entanglement_entropy,
          siteinds,length,div,M_MPS},
    emphstyle=\color{codeviolet},
    numbers=left,
    numberstyle=\tiny\color{linegray},
    stepnumber=1,
    numbersep=8pt,
    frame=single,
    rulecolor=\color{black!25},
    breaklines=true,
    captionpos=b,
    showstringspaces=false,
    tabsize=4
}

\lstdefinestyle{pythonstyle}{
    language=Python,
    backgroundcolor=\color{codegray},
    basicstyle=\ttfamily\footnotesize,
    keywordstyle=\color{codeblue}\bfseries,       
    stringstyle=\color{codered},                  
    commentstyle=\color{codegreen}\itshape,       
    alsoletter=@_,
    morekeywords=[2]{int,float,str,list,dict,set,tuple,bool,object,type,range,enumerate,len,print,abs,sum,min,max,zip,map,filter},
    keywordstyle=[2]\color{codeviolet}\bfseries,
    morekeywords=[3]{@dataclass,@staticmethod,@classmethod,@property,@jit,@njit},
    keywordstyle=[3]\color{codeorange}\bfseries,
    emph={xacc,getAccelerator,qalloc,execute,getInformation,buffer,circuit,cost_function,N,avg_entropy,std_dev},
    emphstyle=\color{codeviolet},
    numbers=left,
    numberstyle=\tiny\color{linegray},
    stepnumber=1,
    numbersep=8pt,
    frame=single,
    rulecolor=\color{black!25},
    breaklines=true,
    captionpos=b,
    showstringspaces=false,
    tabsize=4,
    keepspaces=true,
    sensitive=true
}

\begin{document}

\title{Integrating Julia-ITensors into the Tensor Network Quantum Virtual Machine (TNQVM)}

\author{Zachary W. Windom}
\email{windomzw@ornl.gov}
\orcid{zachincludeyourorcid}
\affiliation{%
\institution{Quantum Information Science Section, Quantum Computational Science Group, Oak Ridge National Laboratory}
\city{Oak Ridge}
\state{Tennessee}
\country{USA}
}

\author{Daniel Claudino}
\email{claudinodc@ornl.gov}
\orcid{0000-0002-8860-0689}
\affiliation{%
\institution{Quantum Information Science Section, Quantum Computational Science Group, Oak Ridge National Laboratory}
\city{Oak Ridge}
\state{Tennessee}
\country{USA}
}

\author{Vicente Leyton-Ortega}
\email{leytonorteva@ornl.gov}
\orcid{0000-0003-4255-9892}
\affiliation{%
\institution{Quantum Information Science Section, Quantum Computational Science Group, Oak Ridge National Laboratory}
\city{Oak Ridge}
\state{Tennessee}
\country{USA}
}
\thanks{This manuscript has been authored by UT-Battelle, LLC, under Contract No.~DE-AC05-00OR22725 with the U.S.~Department of Energy. The United States Government retains and the publisher, by accepting the article for publication, acknowledges that the United States Government retains a non-exclusive, paid-up, irrevocable, world-wide license to publish or reproduce the published form of this manuscript, or allow others to do so, for the United States Government purposes. The Department of Energy will provide public access to these results of federally sponsored research in accordance with the DOE Public Access Plan (https://energy.gov/doe-public-access-plan).}

\begin{abstract}
The Tensor Network Quantum Virtual Machine (TNQVM) is a high-performance classical circuit simulation backend for the eXtreme-scale ACCelerator (XACC) framework that leverages the Intelligent Tensor (ITensor) library for tensor network--based quantum circuit simulation. However, TNQVM's original C++ ITensor backend is tied to an older integrated release, limiting access to newer tensor network algorithms, diagnostics, and performance improvements available in the actively developed Julia-based ITensors ecosystem. We introduce JuliaITensorTNQVM, an interoperability layer that bridges TNQVM's C++ visitor infrastructure and the Julia-ITensors runtime through a C-compatible application binary interface. This design preserves the existing XACC/TNQVM programming model while enabling access to modern tensor network capabilities, including entanglement entropy diagnostics exposed directly to XACC. We evaluate the implementation through two studies: a Page-curve verification protocol using Haar-random states, and QAOA MaxCut simulations on 3-regular graphs. Within these tested regimes, results are consistent with expected entanglement behavior and established scaling trends, supporting JuliaITensorTNQVM as a practical modernization path for tensor network simulation in TNQVM.
\end{abstract}

\maketitle

\section{Introduction}

Quantum computers are expected to offer algorithmic advantages over classical approaches in specific problem domains. However, the boundary separating quantum advantage---whereby the capabilities of a quantum computer surpass those of classical heuristics---is opaque and continues to be an active area of research. As near-term quantum devices remain restricted in qubit count, connectivity, and circuit depth, their practical utility for algorithmic exploration is limited. In this context, classical simulation plays a critical role by providing a controlled environment in which quantum algorithms can be developed, tested, and systematically analyzed prior to deployment on hardware.

To support this development cycle, a flexible quantum software stack must accommodate a diverse range of execution targets, including both physical quantum processors and classical simulators. The XACC quantum--classical computing framework was designed to address this need by providing a universal API for quantum program development, compilation, and execution across heterogeneous backends~\cite{mccaskey2020xacc}. XACC adopts a co-processor programming model that manages the interaction between CPU and QPU resources while remaining agnostic to underlying hardware and programming language choices~\cite{mccaskey2018language}. Its extensible, plugin-based architecture exposes a unified \texttt{Accelerator} interface through which a wide variety of execution targets---both physical QPUs and classical simulation backends---can be accessed without modification to user-level code.

Within this ecosystem, the Tensor Network Quantum Virtual Machine (TNQVM) serves as a backend execution engine that integrates simulation capabilities based on tensor networks (TN) into XACC. TNQVM is designed for TN simulation across a variety of hardware types and execution models, including multicore, GPU, and distributed-memory configurations in heterogeneous and homogeneous HPC systems~\cite{mccaskey2018validating}. Its pluggable interface supports multiple numerical backends, including the Exascale Tensor Networks (ExaTN)~\cite{lyakh2022exatn, nguyen2022} and the Intelligent Tensor (ITensor)~\cite{fishman2022itensor}. However, the ITensor-based backend within TNQVM is tightly coupled to an older C++ ITensor release whose API has not evolved in step with the broader ITensor ecosystem. This coupling has limited TNQVM's ability to incorporate new algorithms, diagnostics, and performance optimizations emerging from the TN community.

A substantial fraction of recent ITensor development has shifted toward a Julia-based implementation~\cite{itensor_julia}. Julia ITensors provides a high-level, expressive programming model while maintaining competitive performance, and many recent advances in TN algorithms in ITensors are implemented first---or exclusively---within the Julia environment. The divergence between the rapidly evolving Julia ITensors ecosystem and TNQVM's reliance on an older C++ backend has created challenges in terms of feature coverage, extensibility, and long-term sustainability. Addressing this gap requires not merely updating a dependency, but rethinking the interface between TNQVM's execution infrastructure and its numerical backend.

Rather than reimplementing TNQVM functionality entirely in Julia or requiring users to adopt a separate simulation workflow, we pursue a strategy that preserves TNQVM's existing integration within the XACC ecosystem. To this end, we introduce JuliaITensorTNQVM, an interoperability layer that connects TNQVM to Julia ITensors through a C-compatible application binary interface (ABI). The key design choice underlying this integration is the use of Julia's native C foreign function interface (\texttt{@ccallable}-annotated functions compiled into a shared library via \texttt{PackageCompiler.jl}), which produces a standard shared library that can be linked directly against TNQVM without requiring a Julia installation at the call site or introducing the overhead of inter-process communication. Memory management responsibilities are cleanly partitioned at this boundary: TNQVM retains ownership of its C++-side objects and passes call arguments across the ABI, while Julia-side allocations---including the MPS state representation and SVD buffers---are managed by the Julia garbage collector and the JuliaITensorTNQVM binding library.

This design preserves the stability of the TNQVM plugin interface to XACC, and consequently the user-facing programming model is unchanged: quantum circuits are constructed, executed, and analyzed through the standard XACC programming interface with no modifications to existing workflows. The only user-visible extension is the availability of new entanglement diagnostics---such as bipartite entanglement entropy---that are computed internally by the Julia ITensors backend and persisted in the XACC buffer for downstream analysis. This approach maintains compatibility with existing workflows, minimizes disruption to downstream users, and provides a sustainable pathway to adopt new TN capabilities as they are developed within the Julia ITensors ecosystem.

We validate JuliaITensorTNQVM through two complementary studies. First, we employ a Page curve verification protocol based on Haar-random states to confirm the correctness of gate application, tensor contraction, canonicalization, and measurement routines across the full simulation stack. Second, we apply the integrated XACC/TNQVM/Julia-ITensors software stack to simulations of the Quantum Approximate Optimization Algorithm (QAOA) for the MaxCut problem on 3-regular graphs, demonstrating that the simulator correctly captures entanglement dynamics and recovers established universal scaling laws relating bond dimension, entanglement entropy, and optimization quality. The main contributions of this work are as follows:

\begin{itemize}
    \item We present JuliaITensorTNQVM, a Julia--C++ interoperability layer that integrates the Julia ITensors ecosystem into TNQVM through a C-compatible ABI, preserving the existing XACC execution model and user-facing interface without modification.
    \item We provide a modernization pathway for TNQVM that preserves existing functionality and supports rapid adoption of newly developed TN algorithms as they emerge within the Julia ITensors ecosystem.
    \item We expose entanglement entropy diagnostics through the standard XACC buffer interface, enabling quantitative analysis of truncation effects and entanglement growth during circuit simulation.
    \item We perform systematic verification and validation, including Page curve analysis for Haar-random circuits and QAOA simulations for MaxCut on 3-regular graphs of varying size, confirming implementation correctness and recovering established scaling laws.
\end{itemize}

The remainder of this paper is organized as follows. Section~\ref{sec:related} situates JuliaITensorTNQVM within the landscape of existing quantum circuit simulators and TN frameworks. Section~\ref{sec:methodology} describes the software architecture, including the TNQVM visitor restructuring, the JuliaITensorTNQVM C API, and the user-facing Python interface, followed by the verification and validation protocols. Section~\ref{sec:results} presents results from the Page curve verification and QAOA MaxCut benchmarks, and Section~\ref{sec:conclusion} concludes with a discussion of the implications and future directions.

\section{Related Works} \label{sec:related}

A wide range of classical quantum circuit simulators have been developed, including state-vector simulators such as Qiskit Aer,\cite{qiskit2024} qsim\cite{quantumsim}, and the NWQ suite.\cite{li2021sv,suh2024simulating} Unfortunately, classical representation of a (general) quantum state requires an exponential amount of memory, which limits simulators of this type to modest system sizes. Extensions based on stabilizer\cite{aaronson2004improved} or Clifford techniques\cite{bennink2017unbiased,bravyi2019simulation,bravyi2016improved,gidney2021stim} can efficiently simulate circuits (almost) exclusively restricted to Clifford gates.

TN methods offer an alternative approach by exploiting limited entanglement structure in quantum states.\cite{reyes2020multi,gray2021hyper,markov2008simulating} Matrix product state (MPS) and projected entangled pair state techniques, for example, have been widely used to simulate low-depth circuits, one-dimensional systems, and variational algorithms.\cite{PhysRevLett.69.2863, PhysRevB.48.10345,verstraete2008matrix,orus2014practical,perez2006matrix,bridgeman2017hand} Software frameworks supporting TN simulation include ITensor (C++ and Julia), ExaTN,\cite{lyakh2022exatn} and QTensor.\cite{lykov2021performance} While these approaches can dramatically extend the reach of classical simulation,\cite{huang2020classical} they rely on truncation strategies that introduce approximation error and require careful control of bond dimension and entanglement growth.
The Julia ITensors library has emerged as a particularly active platform for developing new TN algorithms, offering high-level abstractions, performance portability, and rapid prototyping capabilities.\cite{itensor_julia} Many recent advances in TN simulation are implemented first—or exclusively—in Julia, motivating the need for interoperability with existing simulation infrastructures.

Variational algorithms such as QAOA have been a primary target for TN simulation due to their structured circuits and moderate entanglement growth at low depth, which are often compatible with MPS methods. Previous work has demonstrated large-scale MPS-based QAOA simulations implemented in Julia,\cite{feeney2025mps} but rely heavily on approximations. For example, a large state-vector QAOA simulation on the 3-regular graph with one layer ($p=1$) and 210 qubits required 1,024 nodes of the Theta supercomputer,\cite{lykov2022tensor} illustrating the computational cost of this regime. It is noteworthy that these software developments function as standalone software stacks rather than integrated quantum programming environments. 

The present work differs from prior efforts by emphasizing infrastructure integration rather than standalone simulation. By interfacing Julia-ITensors with TNQVM through JuliaITensorTNQVM, we remove a key architectural bottleneck that had previously constrained the evolution of TNQVM. This integration provides sustained access to modern TN algorithms while ensuring the long-term maintainability of TNQVM and its seamless incorporation into full quantum application workflows. The JuliaITensorTNQVM API combines the flexibility and modern capabilities of the Julia ecosystem with the hardware-agnostic execution model and tooling provided by XACC. As a result, this approach complements existing TN simulators and enables advanced diagnostics and algorithmic studies within a unified quantum software framework.

\section{Methodology} \label{sec:methodology}
Modifications to the original XACC/TNQVM/ITensors pipeline are summarized in the following subsections. We initially introduce how XACC enables creation of the necessary workflow, including its reliance on the \texttt{Accelerator} interface to target desired backends, and summarize how the TNQVM \texttt{Accelerator} concrete implementation has been adapted to accommodate the recent changes associated with the updated Julia-ITensors numerical backend. We then briefly discuss the JuliaITensorTNQVM API and highlight some of its functionality. Finally, we showcase new functionality in a representative QAOA problem, and then discuss how the correctness of our implementation is assessed.    
  
\subsection{XACC and the TNQVM Accelerator}

XACC represents quantum programs as an intermediate representation (IR) tree composed of \texttt{Instruction}s, e.g., gates, which can be gathered as \texttt{CompositeInstruction}s, e.g., circuits, and program control flows. XACC traverses the IR tree by visiting each node via its \texttt{BaseInstructionVisitor} interface, which is sub-typed to the underlying backend targeted for execution. Within TNQVM, this visitor infrastructure is what maps individual gate instructions to concrete tensor operations on whichever numerical backend has been selected. 

\begin{figure}[ht!]
\centering
\includegraphics[width=\columnwidth]{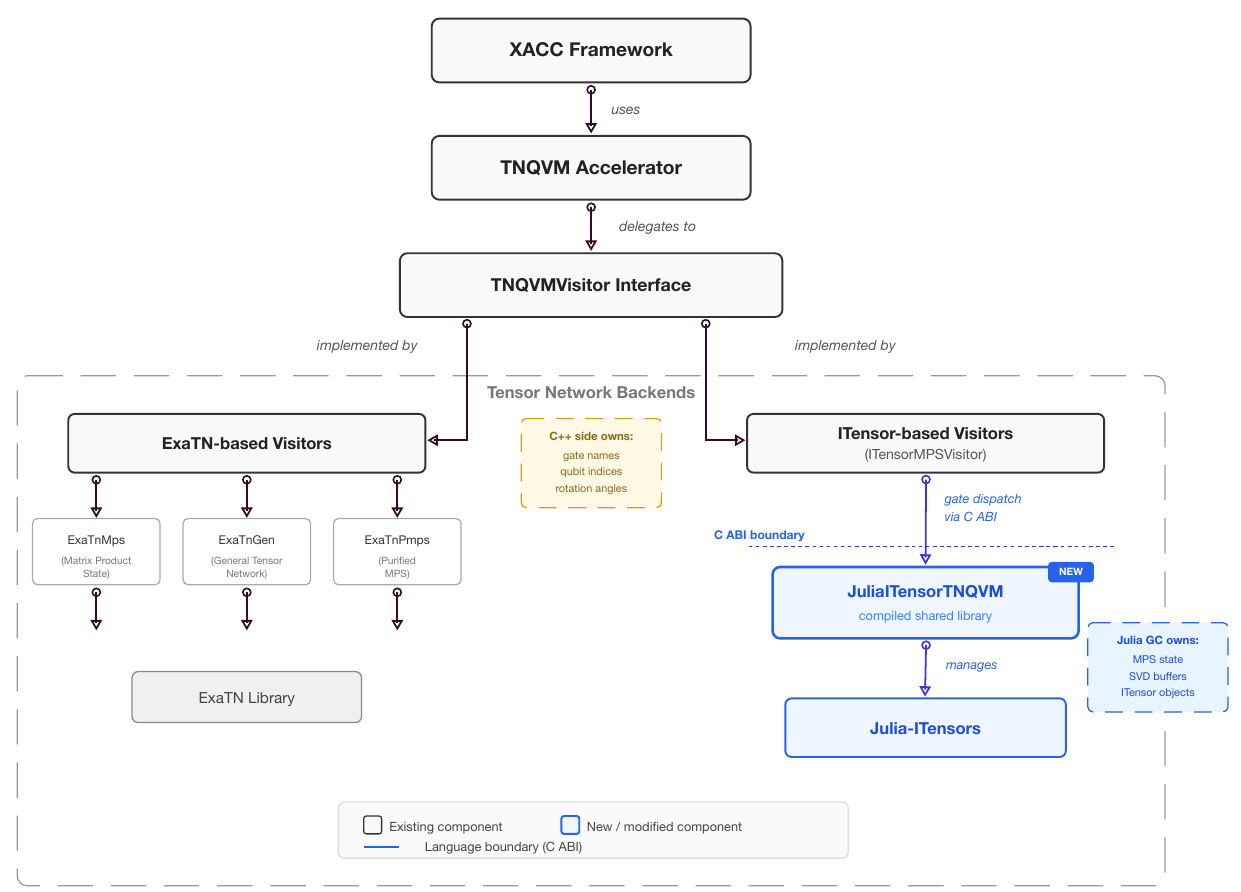} 
\caption{Architecture of the XACC/TNQVM execution pipeline. The XACC framework delegates circuit execution to the TNQVM Accelerator, which dispatches gate operations through the visitor interface to backend-specific implementations. The ITensor-based visitor path (right) crosses a C ABI into the JuliaITensorTNQVM compiled library, which manages the Julia runtime and interfaces with the Julia-ITensors numerical backend. Memory management responsibilities are partitioned at this boundary: the C++ side supplies gate metadata as ABI call arguments, while Julia-side allocations are managed by the Julia garbage collector. The ExaTN-based visitor path (left) and its associated sub-visitors are shown for context.}
\label{fig:pipeline}
\end{figure}

The numerical backend is the component responsible for carrying out the actual tensor operations---constructing and contracting tensors, performing decompositions, and managing the underlying state representation. In the original TNQVM design, this role was filled by an earlier C++ ITensor library integration whose API has since diverged from the actively maintained Julia-based ITensors ecosystem. As a result, TNQVM could not access recent algorithmic improvements, diagnostic capabilities, or performance optimizations developed within Julia-ITensors without a fundamental change to the numerical backend interface.

To be able to leverage the most up-to-date features in ITensors, we restructured the TNQVM visitor to delegate tensor operations to the \texttt{JuliaITensorTNQVM} library, which serves as an intermediary between the C++ visitor infrastructure and the Julia-ITensors runtime. This integration required selecting an interoperability mechanism between C++ and Julia. We chose Julia's native C foreign function interface (\texttt{@ccallable}-annotated functions compiled via \texttt{PackageCompiler.jl}) over alternatives such as \texttt{CxxWrap.jl}~\cite{cxxwrap} or socket-based inter-process communication. The \texttt{@ccallable} approach produces a standard shared library with a C-compatible ABI, which can be linked directly against TNQVM without requiring a Julia installation at the call site and without introducing the overhead or fragility of inter-process communication. This choice yields a clean integration point: TNQVM invokes gate operations, observable evaluations, and state management routines through ordinary C function calls, while the compiled library internally manages the Julia runtime and all Julia-ITensors data structures.

An important consequence of this design is that memory management responsibilities are cleanly partitioned across the language boundary. TNQVM retains ownership of C++-side program objects and supplies gate metadata as ABI call arguments, while Julia-side allocations, including the MPS state representation and SVD buffers, are managed by the Julia garbage collector and the \texttt{JuliaITensorTNQVM} binding library. The Julia runtime itself must be initialized before any tensor operations and must outlive all Julia-allocated objects; \texttt{JuliaITensorTNQVM} exposes dedicated lifecycle management functions (\texttt{initialize}, \texttt{finalize}) to ensure this invariant is maintained. Further details of the Julia-side API are discussed in the following subsection.

The interaction between XACC, TNQVM, and the numerical execution backends is illustrated in Figure~\ref{fig:pipeline}. Listing~\ref{lst:TNQVMsingleqbit} shows how the restructured TNQVM visitor dispatches a single-qubit gate to the \texttt{JuliaITensorTNQVM} library, illustrating the boundary between the C++ visitor logic and the Julia-backed tensor operations.

\begin{lstlisting}[language=C++,
                   mathescape=true,
                   backgroundcolor=\color{black!5},
                   basicstyle=\ttfamily\footnotesize,
    keywordstyle=\color{blue}\ttfamily,
    stringstyle=\color{black}\ttfamily,
    commentstyle=\color{comment-text-color}\ttfamily,
                   caption={Single-qubit gate dispatch in the restructured TNQVM
                   visitor. The visitor extracts gate metadata from the XACC IR
                   and forwards it across the language boundary via the
                   \texttt{JuliaITensorTNQVM} C API.},
                   label={lst:TNQVMsingleqbit}]
void ITensorMPSVisitor::applySingleQubitGate(xacc::Instruction &in_gate) {
  // assert it is a single qubit gate and get name and qubit location
  assert(in_gate.nRequiredBits() == 1);
  auto bit_loc = in_gate.bits()[0] + 1;
  auto gateName = in_gate.name();
  // check if it is a rotation gate
  if (gateName == "Rx" || gateName == "Ry" || gateName == "Rz") {
    double angle = in_gate.getParameter(0).as<double>();
    JuliaITensorTNQVM::applySingleQubitRotGate(bit_loc, gateName.c_str(), angle);
  } else {
    JuliaITensorTNQVM::applySingleQubitGate(bit_loc, gateName.c_str());
  }
}
\end{lstlisting}

\subsection{JuliaITensorTNQVM}

The JuliaITensorTNQVM package consists of three components: the Julia source code implementing TN operations atop Julia-ITensors, a C-compatible API that exposes these operations to external programs, and a precompiled shared library produced via \texttt{PackageCompiler.jl} that is linked against TNQVM at build time. Together, these components enable TNQVM to access the full capabilities of the Julia-ITensors ecosystem through standard C function calls, without requiring a standalone Julia installation or runtime embedding at the user level.

\emph{Gate applications.} Quantum gates are applied to the MPS through a family of C-callable functions that accept gate metadata---qubit index, gate name, and rotation angle where applicable---as ABI call arguments from the TNQVM visitor. Internally, JuliaITensorTNQVM translates this metadata into the appropriate Julia-ITensors operator and applies it to the MPS in-place. For single-qubit gates, the MPS is first orthogonalized to the target site before the gate tensor is contracted. For two-qubit gates, the procedure additionally involves contracting the gate with the two adjacent site tensors, performing an SVD to restore the MPS form, and truncating singular values according to the active bond dimension or cutoff parameters. This sequence---contraction, decomposition, and truncation---is the core computational loop of any MPS-based circuit simulation and is handled entirely by Julia-ITensors. Listing~\ref{lst:JBsingRot} illustrates the structure of a representative gate application function.

\begin{lstlisting}[style=juliastyle,
                   caption={Representative gate application function in JuliaITensorTNQVM. The function receives gate metadata from the C++ visitor, translates it into a Julia-ITensors operator, and applies it to the MPS.},
                   label={lst:JBsingRot}]
Base.@ccallable function applySingleQubitRotGate(bit_loc::Int64, in_gate::Ptr{Cchar}, theta::Float64)::Nothing
    global M_MPS
    orthogonalize!(M_MPS, bit_loc)
    gate_name::String = unsafe_string(in_gate)
    qbit_struct = single_qbit_struct(gate_name, bit_loc, theta)::SingleQubitGate
    M_MPS = apply_gate!(M_MPS, qbit_struct, theta)
    return nothing
end
\end{lstlisting}

\emph{Observable evaluation and entanglement diagnostics.} Beyond gate application, JuliaITensorTNQVM exposes functions for computing observables and entanglement properties of the MPS state. Expectation values are evaluated using the Matrix Product Operator (MPO) functionality provided by Julia-ITensors. Entanglement entropy can be computed at arbitrary bipartitions along the MPS chain: \texttt{computeMidpointMPSEntropy} evaluates the von Neumann entropy at the midpoint (site $N/2$), while \texttt{computeEntanglementEntropy} computes the entropy across a set of bipartition sites, returning both the mean and standard error. The latter is used in the Page curve verification protocol described in Section~\ref{subsec:PagecurveResults}. In all cases, computed quantities are returned across the C ABI boundary to TNQVM and stored in the XACC buffer, making them accessible to downstream analysis through the standard XACC interface. Listing~\ref{lst:JBentropy} shows the half-chain entropy function as an example.

\begin{lstlisting}[style=juliastyle,
    caption={Half-chain entanglement entropy computation exposed through the C API. The result is returned as a C-compatible tuple for transfer across the language boundary.},
    label={lst:JBentropy}]
Base.@ccallable function computeMidpointMPSEntropy()::NTuple{1, Cdouble}
    global M_MPS
    N = length(siteinds(M_MPS))
    half_N = div(N, 2)
    return (compute_entanglement_entropy(M_MPS, half_N), )
end
\end{lstlisting}

This design concentrates all TN logic within the Julia-ITensors ecosystem while presenting a minimal, stable C interface to TNQVM. As new algorithms or diagnostic capabilities are developed within Julia-ITensors---such as improved decomposition strategies, alternative TN geometries, or additional observable types---they can be incorporated into JuliaITensorTNQVM by implementing the corresponding Julia-side logic and exposing it through a new C-callable entry point, without modifications to TNQVM or the broader XACC infrastructure.

\subsection{XACC programming interface}

A central design goal of JuliaITensorTNQVM is that users interact with the Julia-ITensors backend entirely through the existing XACC/TNQVM programming model, with no modifications to circuit construction, compilation, or execution logic. Listing~\ref{lst:Pythonimp} demonstrates this: a TNQVM accelerator instance is obtained from XACC with the Julia-based MPS visitor selected via a single runtime configuration flag (\texttt{"itensor-mps"}). From that point forward, quantum circuits---including parameterized ans\"{a}tze such as QAOA---are constructed, executed, and optimized using the standard XACC interface.

The only user-visible change introduced by the JuliaITensorTNQVM integration is the availability of new diagnostic quantities in the qubit buffer after execution. Specifically, the half-chain entanglement entropy and its statistical variation are computed internally by the Julia-ITensors backend during simulation and persisted in the buffer object. These quantities can be retrieved using the standard \texttt{buffer.getInformation()} interface, as shown in lines 19--20 of Listing~\ref{lst:Pythonimp}. This design ensures that entanglement diagnostics are accessible without requiring users to manage Julia objects, modify their optimization loops, or adopt a separate post-processing workflow.

\clearpage  
\begin{lstlisting}[style=pythonstyle,
    caption={User-facing Python interface to the Julia-ITensors backend. The TNQVM accelerator is configured with a single runtime flag; circuit construction, execution, and optimization proceed through the standard XACC API. Entanglement diagnostics computed by the Julia-ITensors backend are retrieved from the qubit buffer after execution.},
    label={lst:Pythonimp}]
import xacc

# Get instance of TNQVM with the Julia ITensor MPS visitor
acc = xacc.getAccelerator('tnqvm', {"tnqvm-visitor": "itensor-mps"})

# Create graph object with NetworkX and QAOA circuit with XACC
# Create buffer
buffer = qalloc(N)

# Build cost function and run QAOA
def cost_function(x):
    # Bind x to circuit parameters
    # Execute circuit
    acc.execute(buffer, circuit)
    # Calculate cost function value and return
    # return value

# Get entanglement information from the qubit buffer after optimization
avg_entropy = buffer.getInformation("avg-entropy")
std_dev     = buffer.getInformation("stddev-entropy")
\end{lstlisting}

\subsection{Verification and validation of JuliaITensorTNQVM}

\subsubsection{Page curve verification protocol} 
To verify the correctness of JuliaITensorTNQVM's implementation, we require a verification protocol that exercises all critical components of the simulation stack: gate application, tensor contraction, canonicalization, and measurement routines. An effective approach is to simulate highly entangled quantum states and compare computed properties against exact analytical predictions. 

For this purpose, we employ the Page curve verification protocol. \cite{page1993average} It analytically characterizes the entanglement properties of random pure quantum states—specifically, the average entanglement entropy of a subsystem as a function of subsystem size. Random quantum states drawn from the Haar measure (the uniform distribution over all possible pure states) represent maximally generic entangled states, making them ideal test cases: they populate all Schmidt sectors and stress test the simulator's ability to represent and manipulate entanglement.

Given a bipartite Hilbert space $\mathscr{H}=\mathscr{H}^A\otimes \mathscr{H}^B$ where $\text{dim}(\mathscr{H}^A) = 2^{N_A}$, $\text{dim}(\mathscr{H}^B) = 2^{N_B}$, and $N_A \leq N_B$, the average entanglement entropy of subsystem $A$ for a Haar-random pure state is given by
\begin{equation}\label{eq:page_curve}
    S_{\rm Page}(N_A) = \sum_{k=2^{N_B} + 1}^{2^{N}}\frac{1}{k}-\frac{2^{N_A}-1}{2^{N_B+1}} 
\end{equation}
where $N = N_A + N_B$ is the total number of qubits. This analytical formula, now known as the Page curve, provides exact predictions for the entanglement entropy $S_A$ of a subsystem $A$ in a random $N$-qubit pure state. By generating Haar-random states (with $M$ samples) in our simulator and comparing the computed entanglement entropies to Page's analytical results, we can verify that our implementation correctly handles quantum states across the full range of possible entanglement structures. Agreement with the Page curve strongly suggests that gate semantics, tensor index ordering, and measurement protocols are all implemented correctly.

Our verification procedure proceeds as follows: we sample unitaries from the Haar measure, convert them to normalized state vectors of dimension $2^N$, and represent these as MPS with bond dimension $\chi$. For each random state, we compute the entanglement entropy at each position along the MPS chain. At site $i$, the system is bipartitioned into subsystem $A$ (qubits 1 through $i$, giving $N_A = i$) and subsystem $B$ (qubits $i+1$ through $N$, giving $N_B = N - i$). The entanglement entropy at site $i$ is computed from the reduced density matrix of subsystem $A$:
\begin{equation}\label{eq:entropy_computation}
    S_{\chi}(N_A) = - \text{Tr}\left[\rho_A \log(\rho_A)\right]
\end{equation}
where $\rho_A = \text{Tr}_B \left\{ |\psi_{\chi}\rangle\langle\psi_{\chi}| \right\}$ is the reduced density matrix obtained by tracing out subsystem $B$ from the MPS state $|\psi_{\chi}\rangle$ with bond dimension $\chi$. This computation yields an entropy profile $S_{\chi}(N_A)$ across all bipartition sizes $N_A = 1, 2, \ldots, N-1$. A Monte Carlo approach generates statistics over multiple Haar-random samples, allowing us to assess both systematic biases and statistical variations in our implementation.

The verification protocol operates at two levels. First, we verify implementation correctness by comparing the computed entropy profile to the analytical Page curve. When using sufficiently large bond dimension $\chi \geq \chi_{\max} = 2^{N/2}$, the MPS exactly represents the random state. In this regime, the MPS representation is exact up to floating-point precision; any residual deviation of $S_{\chi_{\max}}(N_A)$ from $S_{\rm Page}(N_A)$ arises from finite Monte Carlo sampling (finite $M$) and numerical roundoff rather than truncation. We focus particularly on the center of the chain ($N_A = N/2$), where entanglement is maximal and deviations from theory are most readily detected. Second, we characterize the accuracy-efficiency tradeoffs of MPS approximations by systematically varying the bond dimension $\chi < \chi_{\max}$ and measuring how truncation affects the computed entanglement structure.

To quantify the accuracy of finite-bond-dimension approximations, we define a simulation fidelity metric:
\begin{equation}\label{eq:simulation_fidelity}
    \mathcal{F}_{\rm sim}(\chi) = \frac{S_{\chi}(N/2)}{S_{\rm Page}(N/2)}
\end{equation}
where $S_{\chi}(N/2)$ is the entanglement entropy at the center computed with bond dimension $\chi$, and $S_{\rm Page}(N/2)$ is the analytical Page prediction from Eq.~\ref{eq:page_curve} evaluated at the maximal bipartition ($N_A = N_B = N/2$). A fidelity of $\mathcal{F}_{\rm sim}(\chi) = 1.0$ indicates perfect agreement with the analytical result—achievable when $\chi \geq \chi_{\max}$ allows exact representation of the random state. Values significantly below 1.0 indicate the MPS approximation fails to capture the correct entanglement structure. By varying $\chi$ and measuring how $\mathcal{F}_{\rm sim}(\chi)$ degrades, we characterize both the correctness of our implementation and the accuracy-efficiency tradeoffs inherent in MPS approximations.

This two-level verification strategy—validating correctness against analytical predictions and characterizing approximation quality—provides comprehensive evidence that JuliaITensorTNQVM correctly implements the full MPS simulation pipeline while simultaneously establishing quantitative bounds on the accuracy of finite-bond-dimension approximations.

\subsubsection{Performance validation: QAOA MaxCut benchmarks}

To validate JuliaITensorTNQVM's performance on realistic computational problems, we apply the simulator to the Quantum Approximate Optimization Algorithm (QAOA) for the MaxCut problem on 3-regular graphs. This benchmark serves multiple purposes: it tests the simulator on a well-studied problem with documented behavior, exercises circuits that generate significant volume-law entanglement, and allows us to characterize how MPS bond dimension truncation affects solution quality for practical optimization tasks.

The MaxCut problem seeks to partition graph vertices to maximize the number of edges between partitions. In QAOA, this is encoded in the problem Hamiltonian:
\begin{equation}\label{eq:maxcutH}
    H_P = \frac{1}{2}\sum_{\{i,j\}\in E} w_{ij}(I - Z_iZ_j)
\end{equation}
for a graph $G=(V,E)$ with edges $\{i,j\}\in E$ and edge weights $w_{ij}>0$, where $Z_i$ denotes the Pauli-$Z$ operator applied to qubit $i$. The QAOA algorithm prepares a parameterized quantum state $|\psi(\boldsymbol{\beta}, \boldsymbol{\gamma})\rangle$ through alternating application of problem and mixer unitaries, and variationally minimizes the expectation value $\langle H_P \rangle$ to approximate the ground state energy of this Hamiltonian.

For this validation benchmark, we assess how well the MPS approximation captures the variational optimization landscape by measuring solution quality. Specifically, we define the QAOA approximation quality metric as:
\begin{equation}\label{eq:qaoa_quality}
    \mathcal{F}_{\chi} = \frac{\langle H_P \rangle_{\chi}}{\langle H_P \rangle_{\rm exact}}
\end{equation}
where $\langle H_P \rangle_{\chi}$ is the minimum energy obtained from QAOA simulation using MPS with bond dimension $\chi$, and $\langle H_P \rangle_{\rm exact}$ is the minimum energy obtained using exact state-vector simulation (equivalently, MPS with $\chi_{\rm exact} = 2^{N/2}$). It is important to note that the QAOA state achieving $\langle H_P \rangle_{\rm exact}$ is generally an approximate solution that does not necessarily correspond to the true ground state of the problem Hamiltonian—QAOA is a heuristic algorithm that provides approximate solutions to NP-hard optimization problems. This distinction, along with other nuances of QAOA quality metrics, is discussed further in the Appendix of Ref.~\cite{nakhl2024calibrating} and Ref.~\cite{demange1996approximation}. Our quality metric follows approaches explored in prior work for monitoring MPS approximation reliability in QAOA simulations \cite{dupont2022calibrating}.
In our convention, lower $\langle H_P \rangle$ indicates better performance. Because both numerator and denominator are computed for the same Hamiltonian with the same sign convention and optimizer setup, $\mathcal{F}_{\chi}\approx 1$ indicates close agreement between truncated-MPS and exact-reference simulations. Comparisons of $\mathcal{F}_{\chi}$ across experiments should therefore be interpreted with fixed instance ensembles and optimization settings.

Despite QAOA simulations of the MaxCut problem being NP-hard in general, significant effort has focused on classical circuit simulation \cite{farhi2014quantum,Farhi2014AQA,farhi2016quantum,guerreschi2017practical}. We selected 3-regular graphs as our test case for two reasons: first, the availability of well-documented literature enables comparison with prior results; second, the corresponding system dynamics embed physics that are particularly challenging to capture with MPS. Specifically, entanglement barriers separate the minimally-entangled initial and final states, with entanglement scaling according to a volume law \cite{dupont2022entanglement}. This makes 3-regular MaxCut an ideal stress test for MPS-based simulation, as it probes the limits of bond dimension truncation in the presence of significant entanglement growth.

Exact state-vector simulation scales as $\mathcal{O}(2^N)$ in memory, limiting practical runs to modest $N$. In contrast, an MPS representation scales as $\mathcal{O}(N\chi^2)$ and becomes exact when $\chi=\chi_{\rm exact}=2^{N/2}$. We therefore treat $\chi<\chi_{\rm exact}$ as a controlled approximation and quantify its impact on QAOA MaxCut on 3-regular graphs, which are known to generate substantial volume-law entanglement that stresses low-$\chi$ truncations \cite{dupont2022entanglement,nakhl2024calibrating}.

\paragraph{Simulation Protocol.}
We use the NetworkX \cite{SciPyProceedings_11} Python library to construct 3-regular graphs for a given number of nodes $N$ (equivalently, qubits). For each choice of system size $N$, approximate bond dimension $\chi < \chi_{\rm exact}$, and QAOA circuit depth $p$, we generate 25 randomly constructed 3-regular graphs. Each QAOA optimization is initialized with randomly sampled variational parameters $\beta \in [0,\pi)$ and $\gamma \in [0,2\pi)$ drawn from uniform distributions. We record the minimal cost function expectation value $\langle H_P \rangle$ averaged across all 25 QAOA runs for subsequent analysis, along with reference values obtained from exact simulations ($\chi = \chi_{\rm exact}$) for comparison via Eq.~\ref{eq:qaoa_quality}.

The variational optimization employs the COBYLA algorithm with a maximum of 1000 iterations per optimization run. We assess the QAOA cost function expectation value as a function of circuit depth $p = 1, 2, \ldots, 6$ layers, examining how approximation quality varies with both algorithmic depth and quantum entanglement complexity. Our analysis also examines the impact of varying the MPS bond dimension in powers of two, $\chi = 2^k$ for $k \in \mathbb{N}$, which allows systematic exploration of the accuracy-efficiency tradeoff. Detailed results and discussion of bond dimension scaling are presented in the following subsection.

\section{Results and Discussion}\label{sec:results}

\subsection{Page curve verification results}\label{subsec:PagecurveResults}

We now present results from the Page curve verification protocol to assess the correctness of JuliaITensorTNQVM's implementation. Recall that agreement between computed entanglement entropies and exact Page values indicates correct implementation of gate operations, tensor contractions, canonicalization, and measurements.

\subsubsection{Convergence of Monte Carlo sampling}

\begin{figure}[ht!]
\centering
\includegraphics[width=0.8 \columnwidth]{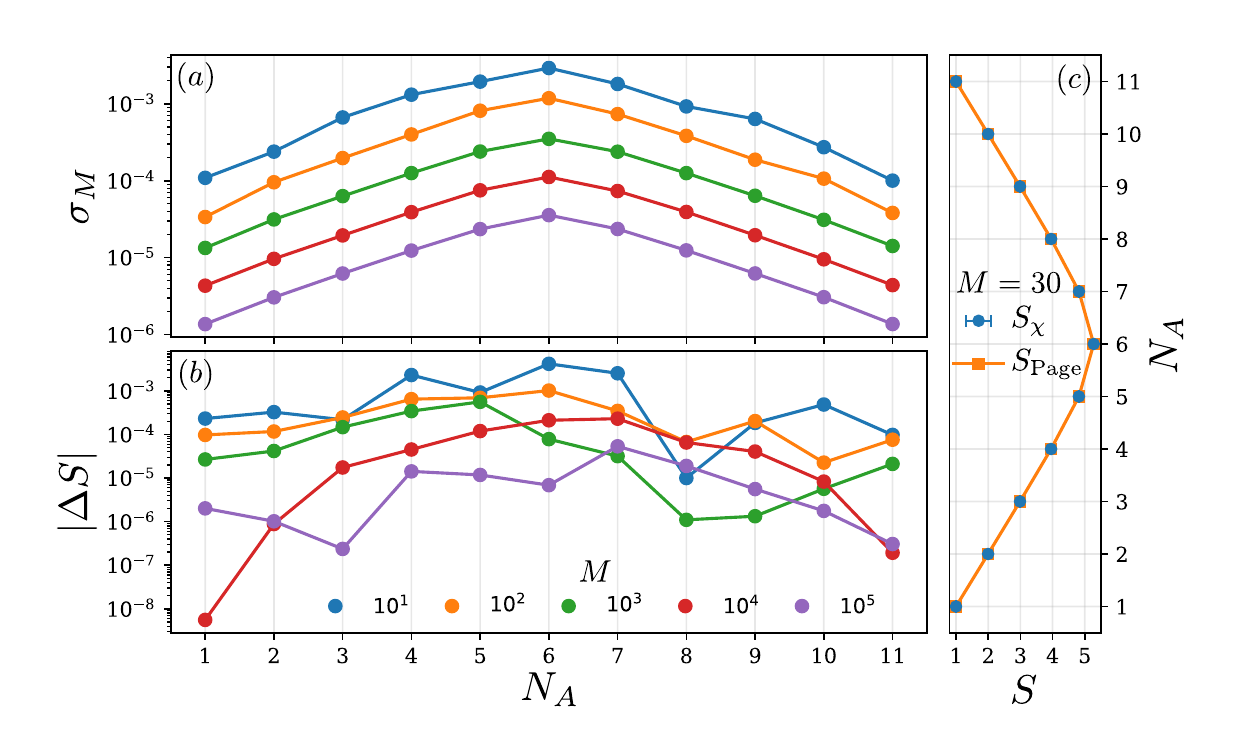} 
\caption{Convergence analysis for the Page curve verification protocol using a 12-qubit MPS with full bond dimension $\chi = 2^6$. (a) Standard error $\sigma_{M}$ as a function of Monte Carlo iterations $M$ for different bipartition sites $N_A$. Error peaks at the center site ($N_A = N/2 = 6$) where entanglement is maximal. (b) Absolute error $|\Delta S| = |S_{\chi}(N_A) - S_{\rm Page}(N_A)|$ between Monte Carlo estimates and exact Page values as a function of subsystem size $N_A$ for different numbers of Monte Carlo iterations $M$. Error decreases systematically with more Monte Carlo samples. (c) Computed entanglement entropy $S_{\chi}$ (blue points) versus exact Page curve prediction $S_{\rm Page}$ (orange line) as a function of subsystem size $N_A$ for $M = 30$ Monte Carlo samples, demonstrating excellent agreement with analytical predictions across all bipartition sizes.}
\label{fig:Error}
\end{figure}

Figure \ref{fig:Error} characterizes the statistical uncertainty in our verification procedure for a 12-qubit system ($N=12$). Panel (a) shows the standard error as a function of Monte Carlo iterations for different bipartition sites, while panel (b) shows the absolute error between Monte Carlo estimates and exact Page values as a function of subsystem size $N_A$. All results use the maximum bond dimension $\chi = 2^6$, which corresponds to an exact state-vector simulation for this system size (since $\chi_{\rm exact} = 2^{N/2} = 2^6$). At this bond dimension, the only source of error arises from finite Monte Carlo sampling of Haar-random unitaries.

As expected, the standard error in panel (a) decreases systematically with increasing Monte Carlo iterations and peaks at the center of the MPS chain (site $N_A = N/2=6$), where entanglement is maximal. The absolute error in panel (b) with respect to exact Page values exhibits similar trends, though with more variability. This variability stems from the presence of both positive and negative deviations that partially cancel when computing unsigned errors. Comparison between both panels suggests that standard error serves as a reasonable proxy for estimating the absolute deviation from Page values. Panel (c) demonstrates that the computed entanglement entropy profile agrees excellently with the analytical Page curve across all bipartition sizes when using $M = 30$ Monte Carlo samples, with $S_{\chi}$ values (blue points) closely tracking the theoretical $S_{\rm Page}$ predictions (orange line).

These convergence studies demonstrate that 10 Monte Carlo cycles achieve standard errors below 0.001, which we adopt for subsequent analysis. This choice is justified by observing that errors from finite Monte Carlo sampling are negligible compared to errors introduced by bond dimension truncation, which we examine next. In the remainder of this section, implementation-correctness claims are therefore tied to agreement with analytical Page-curve predictions within this sampling uncertainty.

\subsubsection{MPS bond dimension truncation effects}

\begin{figure}[ht!]
\centering
\includegraphics[width=0.8 \columnwidth]{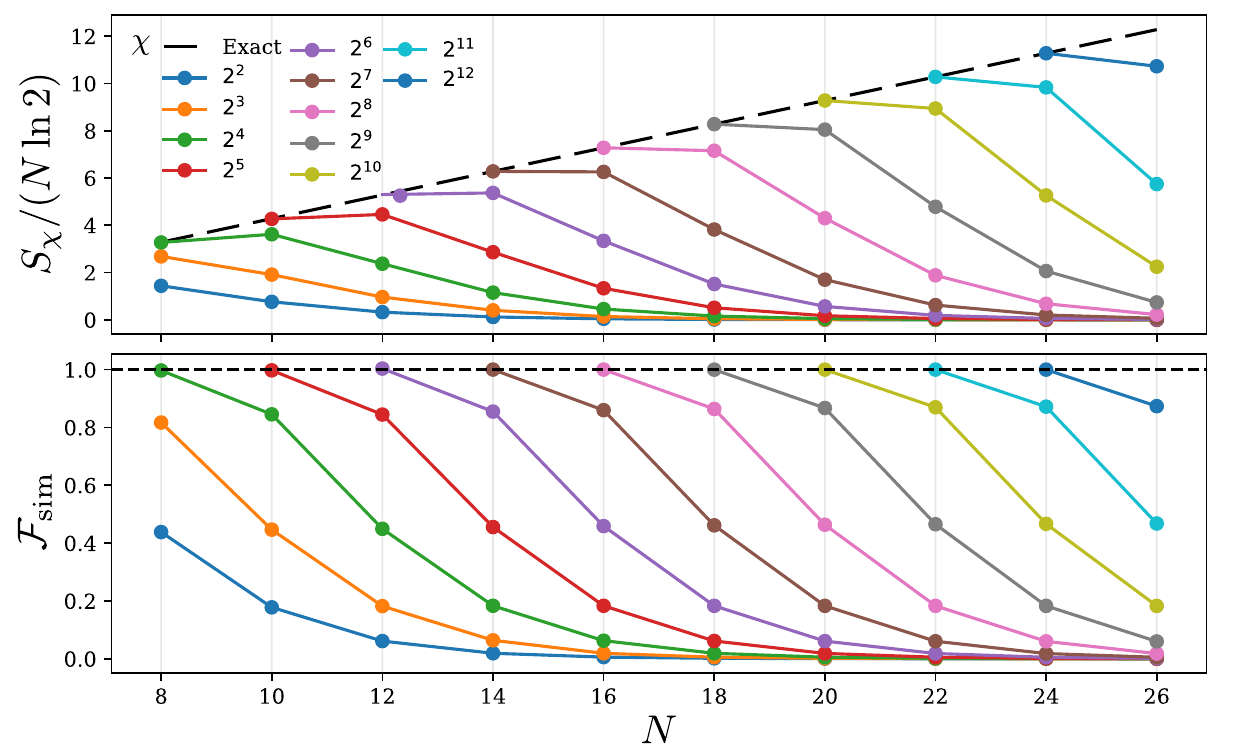} 
\caption{Effect of bond dimension truncation on MPS representation quality for Haar-random states. Top: Normalized entanglement entropy $S_{\chi}/(N\ln 2)$ at the center site ($N/2$) versus system size $N$ for various bond dimensions $\chi$. The exact representation ($\chi = 2^{N/2}$, black dashed line) maintains unit fidelity across all system sizes. Bottom: Simulation fidelity $\mathcal{F}_{\rm sim}(\chi) = S_{\chi}(N/2)/S_{\rm Page}(N/2)$ versus system size $N$. Curves are color-coded by bond dimension. Maintaining constant fidelity across system sizes requires exponential scaling $\chi \propto 2^{N/2}$.}
\label{fig:HaarMPSresults}
\end{figure}

Having validated convergence of the Monte Carlo procedure, we now systematically examine how bond dimension truncation affects the fidelity of MPS representations. Figure \ref{fig:HaarMPSresults} presents two complementary views of this relationship. The top panel shows the normalized entanglement entropy $S_{\chi}/(N\ln 2)$ at the center of the MPS chain (site $N/2$) as a function of system size $N$ for various bond dimensions. The bottom panel shows the corresponding simulation fidelity $\mathcal{F}_{\rm sim}(\chi)$ defined as the ratio of computed entropy to exact Page entropy.

For the exact representation with $\chi = 2^{N/2}$, the normalized entropy agrees with Page values to within numerical precision, yielding fidelity of 1.0 across all system sizes. This agreement validates that JuliaITensorTNQVM correctly implements the quantum operations tested by this protocol: gate application, tensor index ordering, canonicalization procedures, and measurement routines all function as intended.

For truncated bond dimensions, clear patterns emerge. When $\chi = 2^{(N-1)/2}$, the fidelity remains above 0.8 but exhibits a discontinuous drop from the exact result. This drop becomes more gradual as $N$ increases, suggesting the discontinuity may diminish at larger system sizes. However, for $\chi < 2^{(N-1)/2}$, the normalized entropy and fidelity decay exponentially with increasing $N$.

The bottom panel of Figure \ref{fig:HaarMPSresults} reveals a key insight: maintaining constant fidelity across different system sizes requires scaling the bond dimension as $\chi \propto 2^{N/2}$. The curves, color-coded by fixed $\chi$ values, show horizontal alignment when the bond dimension increases by a factor of two for every increment of two qubits. For example, maintaining fidelity slightly above 0.8 requires $\chi = 2^{(N-1)/2}$, while the next lower value $\chi = 2^{(N-2)/2}$ yields fidelity barely above 0.4—a precipitous drop.

This exponential scaling relationship has important implications: for MPS-based quantum simulation of highly entangled states like those in the Haar ensemble, achieving consistent approximation quality demands that bond dimension scales exponentially with system size. When $\chi$ is held constant while $N$ increases, approximation fidelity degrades exponentially, rendering the MPS representation increasingly unreliable.

\subsection{QAOA MaxCut validation results}
Having verified the correctness of JuliaITensorTNQVM's implementation via the Page curve protocol, we now validate the simulator's performance on a realistic computational task: QAOA optimization for the MaxCut problem on 3-regular graphs. This validation addresses two key questions: (1) Can JuliaITensorTNQVM successfully execute QAOA circuits that generate volume-law entanglement? (2) How does bond dimension truncation affect the quality of QAOA optimization results?

\subsubsection{Bond dimension effects on QAOA performance}
\begin{figure}[ht!]
\centering
\includegraphics[width=0.8 \columnwidth]{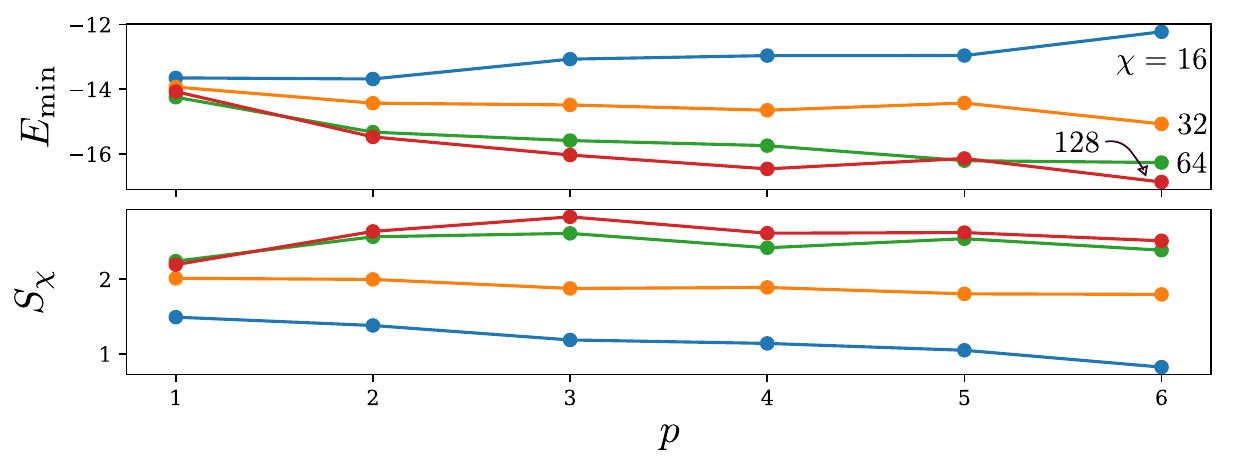} 
\caption{Bond dimension effects on QAOA simulation for 14-qubit MaxCut on 3-regular graphs. Top: Cost function expectation value $\braket{H_P}$ versus QAOA depth $p$. Bottom: Entanglement entropy at the center bipartition (site $N/2=7$) versus QAOA depth. Results are averaged over 25 randomly generated graphs for each bond dimension. Large bond dimensions ($\chi = 64, 128$) show consistent behavior, while smaller values exhibit degraded performance.}
\label{fig:QAOAError}
\end{figure}

Figure \ref{fig:QAOAError} shows how bond dimension $\chi$ impacts two key quantities for a 14-qubit system: the QAOA cost function $\braket{H_P}$ (top panel) and the entanglement entropy at the center of the MPS chain, $\braket{S_{N/2}}$ (bottom panel), both averaged over 25 randomly generated 3-regular graphs. Results are shown as a function of QAOA circuit depth $p=1$ to $6$ layers.

For large bond dimensions ($\chi = 64$ and $\chi = 128$), both the cost function and entanglement entropy exhibit consistent behavior. The entanglement entropy reaches its maximum between $p=1$ and $p=3$ layers, followed by oscillatory decay at larger depths—a signature of the entanglement dynamics in QAOA circuits for MaxCut. The cost function values for $\chi = 64$ and $\chi = 128$ are nearly identical, consistent with our Page curve verification results which showed that $\chi \geq 2^{(N-1)/2} = 2^{6.5} \approx 64$ provides high-fidelity representations for 14-qubit systems.

For smaller bond dimensions ($\chi = 16$ and $\chi = 32$), clear degradation appears. The entanglement entropy curves become flatter and less structured, losing the characteristic peak and oscillatory decay pattern. For $\chi = 16$, the behavior is severely distorted, indicating the MPS representation cannot adequately capture the entanglement generated by QAOA circuits. The $\chi = 32$ results show intermediate behavior: while degraded compared to higher bond dimensions, they retain some qualitative features of the expected dynamics.

These trends align closely with our Page curve verification findings: adequate representation quality requires $\chi \gtrsim 2^{(N-1)/2}$, and approximation fidelity degrades rapidly below this threshold.

To understand how simulation fidelity scales with both system size and bond dimension, we examine universal scaling relationships originally identified by Dupont et al.~\cite{dupont2022calibrating}. Such scaling laws are critical for assessing the classical computational cost required to achieve a target output fidelity and for benchmarking quantum versus classical processing capabilities \cite{dupont2022calibrating}.

\begin{figure}[ht!]
\centering
\includegraphics[width=0.8 \columnwidth]{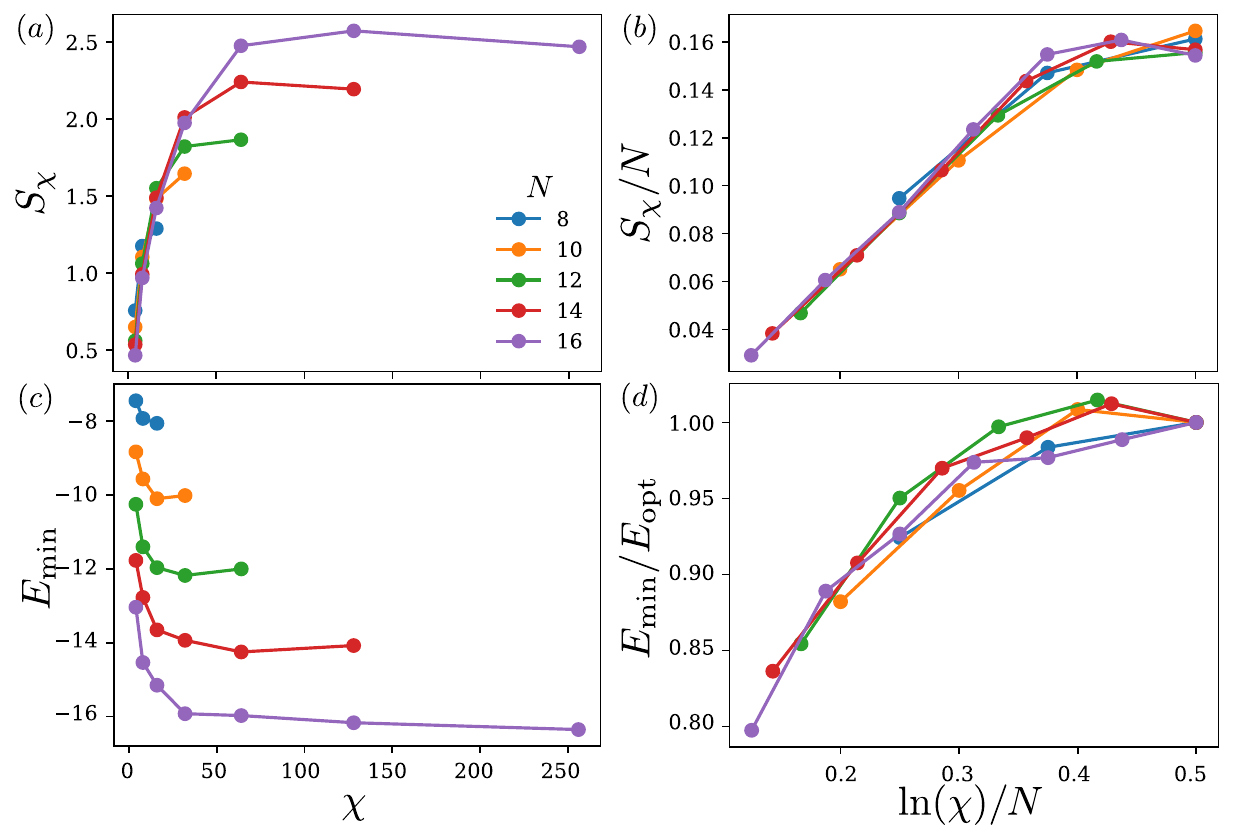} 
\caption{Universal scaling of entanglement entropy and cost function in QAOA MaxCut simulations for system sizes $N = 8, 10, 12, 14, 16$ using single-layer QAOA ($p=1$) on 3-regular graphs. (a) Entanglement entropy at the center bipartition (site $N/2$) versus bond dimension $\chi$. Each system size produces a distinct curve. (b) Collapsed entanglement scaling showing entropy per qubit $S_{\chi}/N$ versus $\ln(\chi)/N$. The collapse is most accurate for $\ln(\chi)/N \lesssim 0.1$. (c) Minimized cost function expectation value $E_{\rm min} \equiv \braket{H_P}$ versus bond dimension $\chi$. (d) Collapsed cost function scaling showing ratio $E_{\rm min}/E_{\rm opt}$ versus $\ln(\chi)/N$, where $E_{\rm opt}$ is the best classical approximation to the minimum energy obtained from the 25-graph ensemble.}
\label{fig:Scalings}
\end{figure}

Figure \ref{fig:Scalings}(a) shows the entanglement entropy at the center bipartition as a function of bond dimension $\chi$ for system sizes $N = 8, 10, 12, 14, 16$ using single-layer QAOA circuits ($p=1$). Each system size produces a distinct curve that asymptotes to different maximum entropy values. The family of curves shows a similar saturation behavior to the Page-curve verification results, with the bond dimension $\chi$ serving as the control parameter.

Remarkably, these distinct curves collapse onto a single universal scaling curve when appropriately rescaled. Figure \ref{fig:Scalings}(b) shows that the entanglement per qubit $S_{\chi}/N$ scales approximately linearly with $\ln(\chi)/N$:
\begin{equation}
\frac{S_{\chi}}{N} \propto \frac{\ln(\chi)}{N}
\end{equation}
This collapse is most accurate in the low-fidelity regime where $S_{\chi}/N = \ln(\chi)/N \lesssim 0.1$, indicating that entanglement generation in QAOA circuits follows predictable scaling behavior across system sizes. The emergence of this universal scaling validates that JuliaITensorTNQVM correctly captures the entanglement dynamics of QAOA, independent of system size.

Figure \ref{fig:Scalings}(c) and (d) examine analogous scaling behavior for the optimization quality, measured by the QAOA cost function. Panel (c) shows the minimized expectation value $\braket{H_P}$ as a function of $\chi$ for different system sizes, while panel (d) plots the ratio $E_{\text{min}}/E_{\text{opt}}$ versus $\ln(\chi)/N$, where $E_{\text{opt}}$ denotes the best classical approximation to the true minimum.

Similar to the entanglement scaling, we observe approximate collapse of the curves when using the rescaled variables. However, the collapse is less ideal in the high-fidelity regime near $E_{\text{min}}/E_{\text{opt}} \rightarrow 1$, where curves show larger spread (approximately 0.05 variation). This discrepancy likely arises from limitations in our classical benchmark: $E_{\text{opt}}$ is estimated from only 25 graph samples using shallow circuits ($p=1$), which may not represent the true optimal solutions. Despite this limitation, the overall scaling behavior confirms that JuliaITensorTNQVM produces QAOA optimization results consistent with established scaling laws, validating its performance on this computational benchmark.

For the tested QAOA MaxCut regime (3-regular graphs, $N=8$--$16$, $p=1$--$6$, and $\chi=16$--$128$), JuliaITensorTNQVM reproduces the expected trend that stable behavior emerges for larger bond dimensions and degrades as $\chi$ is reduced. The observed scaling trends are consistent with prior reports\cite{morris2024performant} and provide practical guidance for selecting bond dimensions in similar variational-simulation settings.

\section{Conclusion}\label{sec:conclusion}

In this work, we presented an interoperability pathway for integrating the Julia-based ITensors ecosystem into TNQVM through \texttt{JuliaITensorTNQVM}. Rather than replacing the existing XACC/TNQVM stack, the approach preserves the visitor-based execution model while enabling access to newer TN capabilities, including entanglement-entropy diagnostics.

The interoperability layer manages the C/C++--Julia boundary through Julia's C foreign function interface, supporting MPS initialization, gate application, and observable evaluation with clearly separated memory-management responsibilities across runtimes.

\textbf{Implementation verification via the Page curve protocol.} For the tested Haar-random benchmarks, computed entanglement entropies are consistent with Page-curve predictions at untruncated bond dimensions (within finite-sampling and numerical error), and they degrade systematically as bond dimension is reduced. This supports correctness of the implemented gate-application, contraction, canonicalization, and measurement pipeline in the evaluated regime.

\textbf{Performance validation via QAOA MaxCut benchmarks.} For 3-regular MaxCut instances with $N=8$--$16$, $p=1$--$6$, and $\chi=16$--$128$, results show the expected dependence of solution quality and entanglement diagnostics on bond dimension. The observed trends are consistent with prior scaling studies and indicate that reduced-$\chi$ approximations degrade predictably in this nonlocal setting.

Taken together, these studies support the utility of JuliaITensorTNQVM for benchmark families that combine highly entangled random states and structured optimization circuits, and they provide practical heuristics for selecting bond dimensions when exact simulation is not feasible.

Our results suggest that while increased bipartite entanglement entropy is not, by itself, a sufficient indicator of improved cost-function minimization in QAOA, it is nevertheless a necessary ingredient for preparing quantum states capable of approaching low-energy solutions for nonlocal optimization problems such as MaxCut on 3-regular graphs. This observation aligns with recent work suggesting that incorporating entanglement-aware heuristics into QAOA may be a promising path forward, particularly in regimes where access to entanglement as a quantum resource is inherently limited \cite{sreedhar2022quantum,chen2022much}.

In summary, JuliaITensorTNQVM extends TNQVM capabilities without disrupting its role within the broader XACC ecosystem. The present results establish a foundation for continued development and broader evaluation as new TN features are incorporated from the Julia ITensors ecosystem.

\section*{Conflict of Interests}
The authors declare that they have no conflicts of interest.

\section*{Author Contributions}
ZW implemented the Julia software components and integrated them with XACC/TNQVM. DC and VLO planned and coordinated the project. All authors contributed to editing the manuscript.

\begin{acks}
This material is based upon work supported by the U.S. Department of Energy, Office of Science. National Quantum Information Science Research Centers, Quantum Science Center.
\end{acks}

\bibliographystyle{ACM-Reference-Format}
\bibliography{references}

\end{document}